\documentclass[12pt]{article}
\usepackage{graphicx}

\begin{document}

\vbox{\vspace{5ex}}

\begin{center}
{\Large \bf Coupled oscillators and Feynman's three papers}

\vspace{2ex}

Y. S. Kim \footnote{electronic address: yskim@physics.umd.edu}\\
Department of Physics, University of Maryland,\\
College Park, Maryland 20742, U.S.A.\\

\end{center}

\vspace{2ex}

\begin{abstract}
According to Richard Feynman, the adventure of our science of
physics is a perpetual attempt to recognize that the different
aspects of nature are really different aspects of the same thing.
It is therefore interesting to combine some, if not all, of
Feynman's papers into one.  The first of his three papers is
on the ``rest of the universe'' contained in his 1972 book on
statistical mechanics.  The second idea is Feynman's parton picture
which he presented in 1969 at the Stony Brook conference on high-energy
physics.  The third idea is contained in the 1971 paper he published
with his students, where they show that the hadronic spectra on Regge
trajectories are manifestations of harmonic-oscillator degeneracies.
In this report, we formulate these three ideas using the mathematics
of two coupled oscillators.  It is shown that the idea of entanglement
is contained in his rest of the universe, and can be extended to a
space-time entanglement.  It is shown also that his parton model
and the static quark model can be combined into one Lorentz-covariant
entity.  Furthermore, Einstein's special relativity, based on the
Lorentz group, can also be formulated within the mathematical
framework of two coupled oscillators.
\end{abstract}

\newpage

\section{Introduction}\label{intro}

According to Feynman, {\it the adventure of our science of physics
is a perpetual attempt to recognize that the different aspects of
nature are really different aspects of the same thing.}

According to what he said above, Feynman is saying or at least trying
to say the same thing in his numerous papers.  Thus, his ultimate goal
was to combine all those into one paper.  Feynman published about 150
papers.  I am not able to combine all those into one paper.  However,
let us see whether we can combine the following three papers he
published during the period 1969-1972.

\begin{itemize}

\item[1.] In 1969, Feynman invented partons by observing hadrons moving with
     velocity close to that of light~\cite{fey69}.  Hadrons are collection
     of partons whose properties are quite different from those of the
     quarks, invented by Gell-Mann.  According to the quark model, hadrons
     are quantum bound states like the hydrogen atom.  While quarks and
     partons appear differently to us, are they the same covariant entity
     in different limiting cases?

     This is a Kantian question which Einstein addressed so brilliantly.
     The energy-momentum relation for slow particles is $E = p^2 /2m $,
     while it is $E = cp$ for fast-moving particles.  Einstein showed that
     they come from the same formula in different limits.  His formula was
     of course $E = \sqrt{(mc^2)^2 + (cp)^2}$.

\item[2.] In his paper on harmonic oscillators~\cite{fkr71}, Feynman notes
     the existence
     of Feynman diagrams for tools of quantum mechanics in the relativistic
     regime.  However, for bound-state problems, he suggests that harmonic
     oscillator could be more effective. Needless to say,     he knew that
     these two methods should solve the same problem of combining quantum
     mechanics and relativity.

     The 1971 paper Feynman wrote with his students contains many original
     ideas~\cite{fkr71}.  However, it is generally agreed that this
     paper is somewhat short in mathematics.  With Marilyn Noz, I
     wrote a book on this subject, and its title is Theory and Applications
     of the Poincar\'e Group.   Our earliest paper on this subject was
     published in 1973.

\item[3.] In his book on statistical mechanics~\cite{fey72}, Feynman divides
    the quantum universe into two systems, namely the world in which we
    do physics, and the rest of the universe beyond our control.
    He thinks we can obtain a better understanding of our observable
    system by considering the universe outside the system.
    If we assume that the same set of physical laws is applicable
    to this rest of the universe, Feynman was talking about two entangled
    systems.

\end{itemize}

    The best way to understand the abstract concept of Feynman's rest
    of the universe is to use two coupled harmonic oscillators,
    where  one is the world in which we do physics, while the other is
    beyond our control and thus is the rest of the universe.  If those
    two oscillators are observed by two different observers, the system
    becomes entangled.

The basic advantage of coupled harmonic oscillators is that the physics is
perfectly transparent thanks to mathematical simplicity.  What is surprising
is that all three of the above subjects can be formulated in terms of
two coupled harmonic oscillators.  In this way, we are able to combine all
three of the above-mentioned research lines into a single physical
problem.

In section~\ref{coupled}, we discuss the quantum mechanics of two
coupled oscillators and write the wave functions in terms of the
variables convenient for studying all three of Feynman's papers
we intend to discuss in this paper.  In section~\ref{restof},
the coupled oscillator system is used to illustrate Feynman's
rest of the universe.  One of the oscillators corresponds to the
world in which we do physics.  The other is in the rest of the
universe.

In section~\ref{feyosc}, we elaborate on Feynman's point that, while
quantum field theory is effective in solving scattering problems
through his Feynman diagrams, it is more convenient to use harmonic
oscillators for bound-state problems.
Paul A. M. Dirac was quite fond of harmonic oscillators.  In
section~\ref{dqm}, we review his efforts to construct relativistic
quantum mechanics using single and couple harmonic
oscillators~\cite{dir27}-\cite{dir63}.

In section~\ref{covham}, it is shown possible to construct a
Lorentz-covariant harmonic oscillators by combining Dirac's work
and the paper Feynman wrote with his students~\cite{fkr71}.
In section~\ref{feydeco}, we discuss in detail Feynman's decoherence
effect contained in his parton picture.

It is known that Einstein in his early years was influenced by
a philosopher named Immanuel Kant~\cite{howard05}.  The same thing
could appear differently to different observers depending on
where they are or how they look at.  If Feynman felt that he was
looking for the same physics while writing different papers,
he was looking for the same thing with different view points.
Feynman was a Kantianist.

In the Appendix, we give a physicist's interpretation of Kantianism,
and we then conclude that, like Einstein, Feynman was a Kantianist.

\section{Coupled Harmonic Oscillators}\label{coupled}

Let us start with two coupled oscillators described by the Hamiltonian
\begin{equation}
H = {1\over 2}\left\{{1\over m} p^{2}_{1} + {1\over m}p^{2}_{2}
+ A x^{2}_{1} + A x^{2}_{2} + 2C x_{1} x_{2} \right\},
\end{equation}
where the oscillators are assumed to have the same mass.
If we choose coordinate variables
\begin{eqnarray}\label{normal}
&{}& z_{1} = {1\over\sqrt{2}}\left(x_{1} + x_{2}\right) , \nonumber\\[2ex]
&{}& z_{2} = {1\over\sqrt{2}}\left(x_{1} - x_{2}\right) ,
\end{eqnarray}
the Hamiltonian can be written as
\begin{equation}
H = {1\over 2m} \left\{p^{2}_{1} + p^{2}_{2} \right\} +
{K\over 2}\left\{e^{-2\eta} z^{2}_{1} + e^{2\eta} z^{2}_{2} \right\} ,
\end{equation}
where
\begin{equation}
  K = \sqrt{A^{2} - C^{2}} ,  \qquad
  \exp(2\eta) = \sqrt{A - C)/(A + C)} .
\end{equation}
The eigenfrequencies are $\omega_{\pm} = \omega e^{{\pm}2\eta}$
with $\omega = \sqrt{K/m}$ .

If $y_{1}$ and $y_{2}$ are measured in units of $(mK)^{1/4} $,
the ground-state wave function of this oscillator system is
\begin{equation}\label{wf01}
\psi_{\eta}(x_{1},x_{2}) = {1 \over \sqrt{\pi}}
\exp{\left\{-{1\over 2}(e^{-2\eta} z^{2}_{1} + e^{2\eta} z^{2}_{2})
\right\} } ,
\end{equation}
The wave function is separable in the $z_{1}$ and $z_{2}$ variables.
However, for the variables $x_{1}$ and $x_{2}$, the story is quite
different, and can be extended to the issue of entanglement.
The key question is how the quantum mechanics in the world of the
$x_{1}$ variable is affected by the $x_{2}$ variable.
If there are two separate measurement processes for these variables,
these two oscillators are entangled.

Let us write the wave function of equation~(\ref{wf01}) in terms of
$x_{1}$ and $x_{2}$, then
\begin{equation}\label{wf02}
\psi_{\eta}(x_{1},x_{2}) = {1 \over \sqrt{\pi}}
\exp\left\{-{1\over 4}\left[e^{-2\eta}(x_{1} + x_{2})^{2} +
e^{2\eta}(x_{1} - x_{2})^{2} \right] \right\} .
\end{equation}
When the system is decoupled with $\eta = 0$, this wave function becomes
\begin{equation}
\psi_{0}(x_{1},x_{2}) = \frac{1}{\sqrt{\pi}}
\exp{\left\{-{1\over 2}(x^{2}_{1} + x^{2}_{2}) \right\}} .
\end{equation}
The system becomes separable and becomes disentangled~\cite{hkn99ajp}.

The wave functions given in this section are well defined in the present
form of quantum mechanics.  These wave functions serve as the basic
scientific language for all three of Feynman's papers we propose to
study in this report.

\section{Feynman's Rest of the Universe}\label{restof}

In his book on statistical mechanics~\cite{fey72}, Feynman makes the
following statement about the density matrix. {\it When we solve a
quantum-mechanical problem, what we really do is divide the universe
into two parts - the system in which we are interested and the rest
of the universe.  We then usually act as if the system in which we
are interested comprised the entire universe.  To motivate the use
of density matrices, let us see what happens when we include the part
of the universe outside the system}.

Feynman then wrote a formula
\begin{equation}\label{wff1}
\psi(x, y) = \sum^{}_{k} C_{k}(y) \phi_{k}(x) ,
\end{equation}
which is a complete-set expansion of the wave function $\phi$ in
the orthonormal set of $\phi_{k}(x),$ but its expansion coefficients
$C_{k}(y)$ depends on another variable $y$.  According to Feynman,
the variable $x$ is for the world in which we do physics, $y$ is
in the rest of the universe where physics or non-physics is done
by a different physicist or a different creature.  In this way,
Feynman introduced the concept of entanglement.

We can use the coupled oscillators to study Feynman's rest of
the universe~\cite{hkn99ajp}.  In order to accommodate Feynman's
original idea more precisely, let us replace $x_1$ and $x_2$ in
equation~(\ref{wf02}) by $x$ and $y$ respectively, and write
the wave function as

\begin{equation}\label{wff2}
\psi_{\eta}(x, y) = {1 \over \sqrt{\pi}}
\exp\left\{-{1\over 4}\left[e^{-2\eta}(x + y)^{2} +
e^{2\eta}(x - y)^{2} \right] \right\} .
\end{equation}
As was discussed in the literature for several different
purposes~\cite{kno79ajp,knp86,knp91}, the oscillator wave
function of equation~(\ref{wf02}) can be expanded as
\begin{equation}\label{expan1}
\psi_{\eta }(x, y) = {1 \over \cosh\eta}\sum^{}_{k}
(\tanh\eta )^{k} \phi_{k}(y) \phi_{k}(x) ,
\end{equation}
where $\phi_{k}(x)$ is the normalized harmonic oscillator wave
function for the $k-th$ excited state.  The coefficient $C_k(y)$
in equation~(\ref{wff1}) now takes the form
\begin{equation}
 C_{k}(y) = \left[\frac{(\tanh\eta)^k}{\cosh\eta}\right] \phi_k(y) .
\end{equation}


\begin{figure}[thb]

\centerline{\includegraphics[scale=0.4]{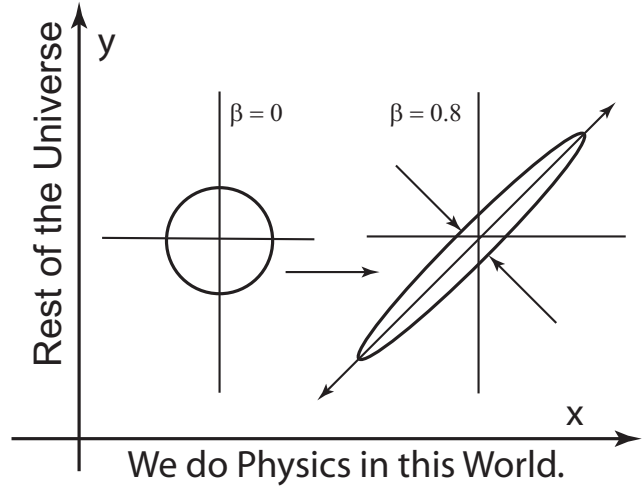}}
\caption{Measurable and non-measurable variables.  Here, the $x$ and $y$
variables are the measurable and non-measurable variables.  We use
$\beta$ for $\tanh\eta$.}\label{ellipse5}

\end{figure}


We can use the coupled harmonic oscillators to illustrate what Feynman
says in his book.  Here we can use respectively $x$ and $y$ for the
variable we observe and the variable in the rest of the universe.  By using
the rest of the universe, Feynman does not rule out the possibility of
other creatures measuring the $y$ variable in their part of the universe.

Using the wave function $\psi_{\eta}(x,y)$ of equation~(\ref{wff1}),
we can construct the pure-state density matrix
\begin{equation}
\rho(x,y;x',y')
= \psi_{\eta}(x,y)\psi_{\eta}(x',y') ,
\end{equation}
which satisfies the condition $\rho^{2} = \rho $:
\begin{equation}
\rho(x,y;x',y') =
\int \rho(x,y;z'',y'')
\rho(z'',y'';x',y') dx'' dy'' .
\end{equation}
If we are not able to make observations on $y$, we should
take the trace of the $\rho$ matrix with respect to the $t$
variable.  Then the resulting density matrix is
\begin{equation}\label{integ}
\rho(x, x') = \int \rho (x,y;x',y) dy .
\end{equation}

The above density matrix can also be calculated from the expansion of
the wave function given in equation~(\ref{expan1}).  If we perform the
integral of equation~(\ref{integ}), the result is
\begin{equation}\label{dmat}
\rho(x,x') = \left({1 \over \cosh(\eta)}\right)^{2}
\sum^{}_{k} (\tanh\eta)^{2k}
\phi_{k}(x)\phi^{*}_{k}(x') .
\end{equation}
The trace of this density matrix is $1$.  It is also straightforward to
compute the integral for $Tr(\rho^{2})$.  The calculation leads to
\begin{equation}
Tr\left(\rho^{2} \right)
= \left({1 \over \cosh(\eta)}\right)^{4}
\sum^{}_{k} (\tanh\eta)^{4k} .
\end{equation}
The sum of this series is $1/\cosh(2\eta)$ which is less than one.

This is of course due to the fact that we are averaging over the $y$
variable which we do not measure.  The standard way to measure this
ignorance is to calculate the entropy defined as
\begin{equation}
S = - Tr\left(\rho \ln(\rho) \right) ,
\end{equation}
where $S$ is measured in units of Boltzmann's constant.  If we use the
density matrix given in equation (\ref{dmat}), the entropy becomes
\begin{equation}
S = 2 \left\{\cosh^{2}\eta \ln(\cosh\eta) -
              \sinh^{2}\eta \ln(\sinh\eta) \right\} .
\end{equation}
This expression can be translated into a more familiar form if
we use the notation
\begin{equation}
\tanh\eta = \exp\left(-{\hbar\omega \over kT}\right) ,
\end{equation}
where $\omega$ is the unit of energy spacing, and $k$ and $T$ are
Boltzmann's constant and absolute Temperature respectively.  The
ratio $\hbar\omega/kT$ is a dimensionless variable.  In terms of
this variable, the entropy takes the form~\cite{hkn89pl,kiwi90pl}
\begin{equation}
S = \left({\hbar\omega \over kT}\right)
\frac{1}{\exp(\hbar\omega/kT) - 1}
- \ln\left[1 - \exp(-\hbar\omega/kT)\right] .
\end{equation}
This familiar expression is for the entropy of an oscillator state
in thermal equilibrium.  Thus, for this oscillator system, we can
relate our ignorance of the time-separation variable to the temperature.
It is interesting to note that the boost parameter or coupling
strength measured by $\eta$ can be related to a temperature variable.

\section{Feynman's Oscillators }\label{feyosc}
\begin{figure}
\centerline{\includegraphics[scale=0.6]{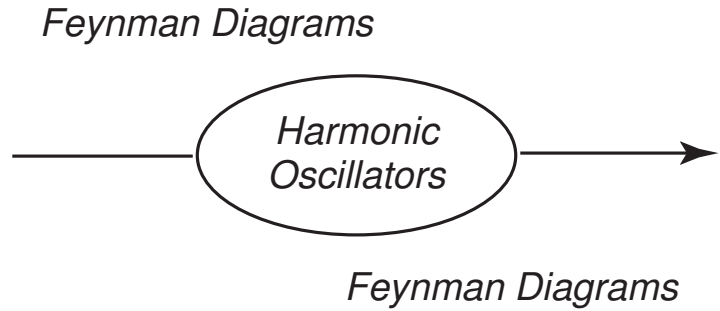}}
\vspace{5mm}
\caption{Feynman's roadmap for combining quantum mechanics with special
relativity.  Feynman diagrams work for running waves, and they provide
a satisfactory resolution for scattering states in Einstein's world.
For standing waves trapped inside an extended hadron, Feynman suggested
harmonic oscillators as the first step.}\label{dff33}
\end{figure}

In his invited talk at the 1970 spring meeting of the American Physical
Society held in Washington, DC (U.S.A.), Feynman was discussing hadronic
mass spectra and a possible covariant formulation of harmonic oscillators.
He noted that the mass spectra are consistent with degeneracy of
three-dimensional harmonic oscillators.  Furthermore, Feynman stressed
that Feynman diagrams are not necessarily suitable for relativistic bound
states and that we should try harmonic oscillators.  Feynman's point was
that, while plane-wave approximations in terms of Feynman diagrams work
well for relativistic scattering problems, they are not applicable to
bound-state problems.  We can summarize what Feynman said in
figure~\ref{dff33}.

In their 1971 paper~\cite{fkr71}, Feynman, Kislinger and Ravndal started
their harmonic oscillator formalism by defining coordinate variables for
the quarks confined within a hadron.  Let us use the simplest hadron
consisting of two quarks bound together with an attractive force, and
consider their space-time positions $x_{a}$ and $x_{b}$, and use the
variables
\begin{equation}
X = (x_{a} + x_{b})/2 , \qquad x = (x_{a} - x_{b})/2\sqrt{2} .
\end{equation}
The four-vector $X$ specifies where the hadron is located in space and
time, while the variable $x$ measures the space-time separation
between the quarks.  According to Einstein, this space-time separation
contains a time-like component which actively participates as in
equation~(\ref{boostm}), if the hadron is boosted along the $z$ direction.
This boost can be conveniently described by the light-cone variables
defined in Eq(\ref{lcvari}).

What do Feynman {\it et al.} say about this oscillator wave function?
In their classic 1971 paper~\cite{fkr71}, they start with the following
Lorentz-invariant differential equation.
\begin{equation}\label{osceq}
{1\over 2} \left\{x^{2}_{\mu} -
{\partial^{2} \over \partial x_{\mu }^{2}}
\right\} \psi(x) = \lambda \psi(x) .
\end{equation}
This partial differential equation has many different solutions depending
on the choice of separable variables and boundary conditions.  This
differential equation gives a complete set of three-dimensional oscillator
solutions with which we are familiar in non-relativistic quantum mechanics.

Indeed, Feynman {\it et al.} studied in detail the degeneracy of the
three-dimensional harmonic oscillators, and compared their results
with the observed experimental data for hadrons in the quark model.
Their work is complete and thorough, and is consistent with the
$O(3)$-like symmetry dictated by Wigner's little group for massive
particles~\cite{wig39}.

Although this paper contained the above mentioned original ideas of
Feynman, it contains some serious mathematical flaws. Feynman {\it et al.}
start with a Lorentz-invariant differential equation for the harmonic
oscillator for the quarks bound together inside a hadron.
For the two-quark system, they write the wave function of the form
\begin{equation}
\exp{\left\{-{1 \over 2}\left(z^2 - t^2 \right) \right\}} ,
\end{equation}
where $z$ and $t$ are the longitudinal and time-like separations between the
quarks.  This form is invariant under the boost, but is not normalizable in
the $t$ variable.  We do not know what physical interpretation to give to
this the above expression.

Yet, Feynman {\it et al.} make an apology that the symmetry is
not $O(3,1)$.  This unnecessary apology causes a confusion not
only to the readers but also to the authors themselves, and makes
the paper difficult to read.  Let us see how we can clear up this
confusion by looking at what Dirac did with harmonic oscillators.

\section{Dirac's Relativistic Quantum Mechanics}\label{dqm}

I was fortunate enough to have private conversations with Paul
A. M. Dirac.  In 1962, when he was visiting the University of
Maryland, I was a first-year assistant professor, and I had to
provide personal convenience to him.  I naturally had an occasion
to ask him what the most problem in physics was at that time,
while Physical Review Letters was carrying articles about Regge
poles, bootstraps, and other S-matrix items.

Dirac was telling me that physicists in general do not understand
the difference between Lorentz invariance and Lorentz covariance.
We said further that we should have a deeper understanding of the
covariance if we are to make progress in physics.  According to
his paper published in 1963~\cite{dir63}, Dirac in 1962 was working
on constructing a representation of the $O(3,2)$ group using two
coupled oscillators.  Since the $S(3,2)$ deSitter group contains
the $O(3,1)$ Lorentz group as a subgroup, Dirac was essentially
telling me to do what I am reporting in this paper.

In 1978, I was again able to talk with him while attending one of
the Coral Gables conferences held in Miami (Florida).  At that time,
I was studying his 1949 paper on ``Forms of Relativistic Dynamics.''
In his ``instant form,'' he writes down a formula which could be
interpreted as a suppression of time-like oscillations.  I asked
him whether I could interpret it in my way.  His reply was that
it depends on how I build the model.

I then pointed out his 1927 paper~\cite{dir27} on the time-energy
uncertainty relation and asked him whether I could use his idea to
suppress time-like oscillations.  Dirac was clearly aware of this paper and
mentioned the word ``c-number'' time-energy uncertainty relation.
In consideration of his age, I did not press him any further.  Let
us see how we could construct a model still within Dirac's framework.

During World War II, Dirac was looking into the possibility of
constructing representations of the Lorentz group using harmonic
oscillator wave functions~\cite{dir45}.  The Lorentz group is the
language of special relativity, and the present form of quantum
mechanics starts with harmonic oscillators.  Presumably, therefore,
he was interested in making quantum mechanics Lorentz-covariant by
constructing representations of the Lorentz group using harmonic
oscillators.

In his 1945 paper~\cite{dir45}, Dirac considered the Gaussian form
\begin{equation}
\exp\left\{- {1 \over 2}\left(x^2 + y^2 + z^2 + t^2\right)\right\} .
\end{equation}
This Gaussian form is in the $(x,~y,~z,~t)$
coordinate variables.  Thus, if we consider Lorentz boost along the
$z$ direction, we can drop the $x$ and $y$ variables, and write the
above equation as
\begin{equation}\label{ground}
\exp\left\{- {1 \over 2}\left(z^2 + t^2\right)\right\} .
\end{equation}
This is a strange expression for those who believe in Lorentz invariance.
The expression $\left(z^2 + t^2\right)$ is not invariant under Lorentz
boost.  Therefore  Dirac's Gaussian form of equation~(\ref{ground}) is not
a Lorentz-invariant expression.

On the other hand, this expression is consistent with his earlier papers
on the time-energy uncertainty relation~\cite{dir27}.  In those papers,
Dirac observed that there is a time-energy uncertainty relation, while
there are no excitations along the time axis.  He called this the
``c-number time-energy uncertainty'' relation.

In 1949, the Reviews of Modern Physics published a special issue to
celebrate Einstein's 70th birthday.  This issue contains Dirac's paper
entitled ``Forms of Relativistic Dynamics''~\cite{dir49}.
In this paper, he introduced his light-cone coordinate system,
in which a Lorentz boost becomes a squeeze transformation.

\begin{figure}[thb]
\centerline{\includegraphics[scale=0.4]{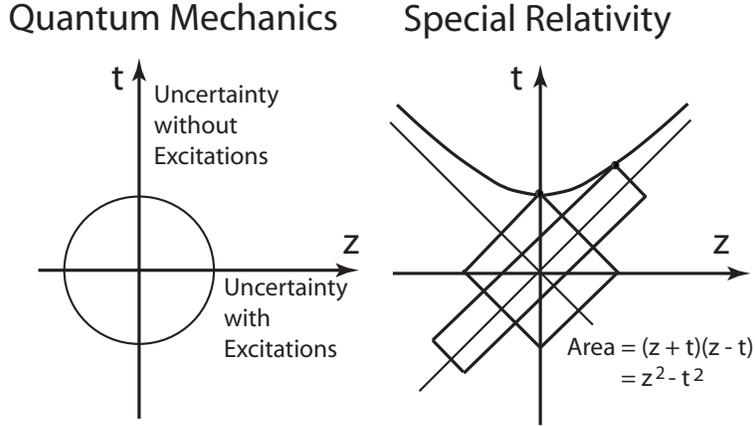}}
\vspace{1mm}
\caption{Dirac's form of relativistic quantum mechanics.}\label{diracq}
\end{figure}

When the system is boosted along the $z$ direction, the transformation
takes the form
\begin{equation}\label{boostm}
\pmatrix{z' \cr t'} = \pmatrix{\cosh\eta & \sinh\eta \cr
\sinh\eta & \cosh\eta } \pmatrix{z \cr t} .
\end{equation}
The light-cone variables are defined as~\cite{dir49}
\begin{equation}\label{lcvari}
u = (z + t)/\sqrt{2} , \qquad v = (z - t)/\sqrt{2} ,
\end{equation}
the boost transformation of equation~(\ref{boostm}) takes the form
\begin{equation}\label{lorensq}
u' = e^{\eta } u , \qquad v' = e^{-\eta } v .
\end{equation}
The $u$ variable becomes expanded while the $v$ variable becomes
contracted, as is illustrated in figure~\ref{diracq}.  Their product
\begin{equation}
uv = {1 \over 2}(z + t)(z - t) = {1 \over 2}\left(z^2 - t^2\right)
\end{equation}
remains invariant.  In Dirac's picture, the Lorentz boost is a
squeeze transformation.

\begin{figure}[thb]

\centerline{\includegraphics[scale=0.4]{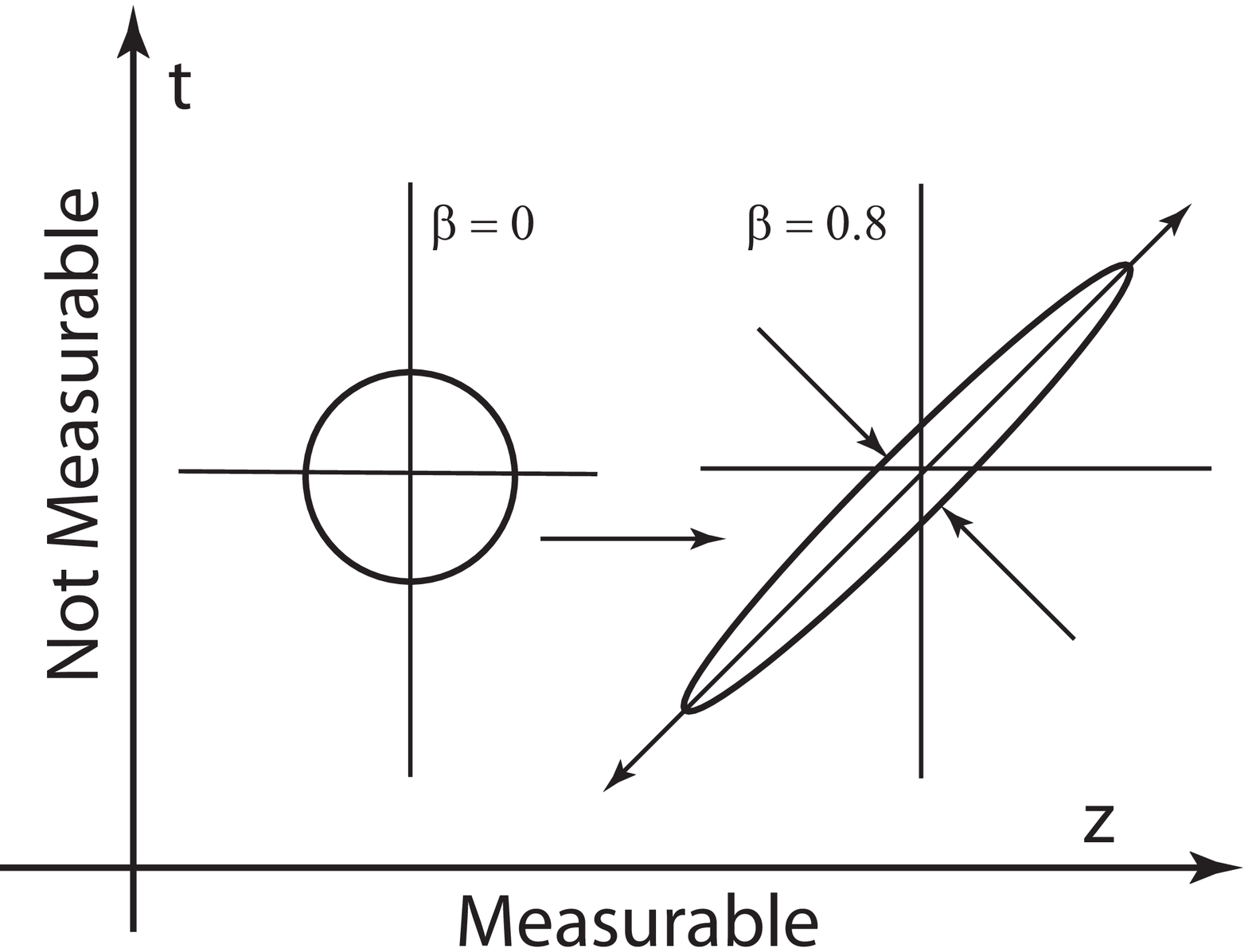}}
\caption{The world in which we do physics and the rest of the
universe.  This figure is identical to that of figure~\ref{ellipse5},
but the axis labels are different, indicating the physics is different.
Yet, the physics of coupled oscillators could be translated into
the physics of Lorentzian space and time.}\label{ellipse3}

\end{figure}

This transformation changes the Gaussian form
of equation~(\ref{ground}) into
\begin{equation}\label{eta}
\psi_{\eta }(z,t) = \left({1 \over \pi }\right)^{1/2}
\exp\left\{-{1\over 2}\left(e^{-2\eta }u^{2} +
e^{2\eta}v^{2}\right)\right\} ,
\end{equation}
as illustrated in figure~\ref{ellipse3}


Let us go back to section~\ref{coupled} on the coupled oscillators.  The
above expression is the same as equation~(\ref{wf01}).  The $x_{1}$ variable
now becames the longitudinal variable $z$, and $x_{2}$ became the time-like
variable $t$.

We can use the coupled harmonic oscillator as the starting point of
relativistic quantum mechanics.  This allows us to translate the quantum
mechanics of two coupled oscillators defined over the space of $x_{1}$
and $x_{2}$ into quantum mechanics defined over the space-time
region of $z$ and $t$.

This form becomes (\ref{ground}) when $\eta$ becomes zero.  The
transition from equation~(\ref{ground}) to equation~(\ref{eta}) is
a squeeze transformation.  It is now possible to combine what Dirac
observed into a covariant formulation of the harmonic oscillator system.
First, we can combine his c-number time-energy uncertainty relation
described in figure~\ref{diracq} and his light-cone coordinate system
of the same figure into a picture of covariant space-time localization
given in figure~\ref{ellipse3}.

The wave function of equation~(\ref{ground}) is distributed within a
circular region in the $u v$ plane, and thus in the $z t$ plane.
On the other hand, the wave function of equation~(\ref{eta}) is
distributed in an elliptic region with the light-cone axes as the major
and minor axes respectively.  If $\eta$ becomes very large, the wave
function becomes concentrated along one of the light-cone axes.  Indeed,
the form given in equation~(\ref{eta}) is a Lorentz-squeezed wave
function.  This squeeze mechanism is illustrated in figure~\ref{ellipse3}.

There are two homework problems which Dirac left us to solve. First, in
defining the $t$ variable for the Gaussian form of equation~(\ref{ground}),
Dirac did not specify the physics of this variable.  If it is going to
be the calendar time, this form vanishes in the remote past and remote
future.  We are not dealing with this kind of object in physics.  What
is then the physics of this time-like $t$ variable?

The Schr\"odinger quantum mechanics of the hydrogen atom deals with
localized probability distribution.  Indeed, the localization condition
leads to the discrete energy spectrum.  Here, the uncertainty relation
is stated in terms of the spatial separation between the proton and
the electron.  If we believe in Lorentz covariance, there must also
be a time-separation between the two constituent particles, and an
uncertainty relation applicable to this separation variable.  Dirac
did not say in his papers of 1927 and 1945, but Dirac's ``t'' variable
is applicable to this time-separation variable.  This time-separation
variable will be discussed in detail in section~\ref{feyosc} for the
case of relativistic extended particles.

Second, as for the time-energy uncertainty relation,  Dirac'c
concern was how the c-number time-energy uncertainty relation without
excitations can be combined with uncertainties in the position space
with excitations.  How can this space-time asymmetry be consistent
with the space-time symmetry of special relativity?

Both of these questions can be answered in terms of the space-time
symmetry of bound states in the Lorentz-covariant regime~\cite{knp86}.
In his 1939 paper~\cite{wig39}, Wigner worked out internal space-time
symmetries of relativistic particles.  He approached the problem by
constructing the maximal subgroup of the Lorentz group whose
transformations leave the given four-momentum invariant.  As a
consequence, the internal symmetry of a massive particle is like the
three-dimensional rotation group which does not require transformation
into time-like space.

If we extend Wigner's concept to relativistic bound states, the
space-time asymmetry which Dirac observed in 1927 is quite consistent
with Einstein's Lorentz covariance. Indeed, Dirac's time variable
can be treated separately.  Furthermore, it is possible to construct
a representations of Wigner's little group for massive
particles~\cite{knp86}.  As for the time-separation which can be
linearly mixed with space-separation variables when the system is
Lorentz-boosted, it has its role in internal space-time symmetry.

Dirac's interest in harmonic oscillators did not stop with his 1945
paper on the representations of the Lorentz group.  In his 1963
paper~\cite{dir63}, he constructed a representation of the $O(3,2)$
deSitter group using two coupled harmonic oscillators.  This paper
contains not only the mathematics of combining special relativity
with the quantum mechanics of quarks inside hadrons, but also forms
the foundations of two-mode squeezed states which are so essential
to modern quantum optics~\cite{knp91,vourdas88,dodo00}.  Dirac did
not know this when he was writing this 1963 paper.  Furthermore,
the $O(3,2)$ deSitter group contains the Lorentz group $O(3,1)$ as a
subgroup.  Thus, Dirac's oscillator representation of the deSitter
group essentially contains all the mathematical ingredients of what
we are studying in this paper.

It is also interesting to note that, in addition to  Dirac and Feynman,
there are other authors who attempted to construct normalizable
harmonic oscillators which can be Lorentz-boosted~\cite{yuka53}.
If the concept of wave functions is to be consistent with Lorentz
covariance, the first wave function has to be the harmonic-oscillator
wave function.

\section{Covariant Oscillators and Entangled Oscillators}\label{covham}

The simplest solution to the differential equation of equation~(\ref{osceq})
takes the form of equation~(\ref{ground}).  If we allow excitations along
the longitudinal coordinate and forbid excitations along the time
coordinate, the wave function takes the form
\begin{equation}\label{wf1}
\psi^{n}_{0} (z,t) = C_{n} H_{n}(z)
\exp{\left\{- {1 \over 2}\left(z^2 + t^2\right)\right\}} ,
\end{equation}
where $H_{n}$ is the Hermite polynomial of the n-th order, and $C_{n}$
is the normalization constant.

If the system is boosted along the z direction, the $z$ and $t$ variables
should be replaced by $z'$ and $t'$ respectively
with
\begin{equation}
z' = (\cosh\eta) z - (\sinh\eta) t , \qquad
t' = (\cosh\eta) t - (\sinh\eta) z .
\end{equation}
The Lorentz-boosted wave function takes the form
\begin{equation}\label{wf2}
\psi^{n}_{\eta} (z,t) = H_{n}(z')
\exp\left\{- {1 \over 2}\left(z'^2 + t'^2\right)\right\} ,
\end{equation}

It is indeed possible to construct the representation of Wigner's
$O(3)$-like little group for massive particles using these oscillator
solutions~\cite{knp86}.  This allows us to use this oscillator system
for wave functions in the Lorentz-covariant world.

However, presently, we are interested in space-time localizations of
the wave function dictated by the Gaussian factor of the ground-state
wave function.  In the light-cone coordinate system, the Lorentz-boosted
wave function becomes
\begin{equation}\label{eta2}
\psi_{\eta }(z,t) = \left({1 \over \pi }\right)^{1/2}
\exp\left\{-{1\over 2}\left(e^{-2\eta }u^{2} +
e^{2\eta}v^{2}\right)\right\} ,
\end{equation}
as given in equation~(\ref{eta}).
This wave function can  be written as
\begin{equation}\label{wf03}
\psi_{\eta }(z,t) = \left({1 \over \pi }\right)^{1/2}
\exp\left\{-{1\over 4}\left[e^{-2\eta }(z + t)^{2} +
e^{2\eta}(z - t)^{2}\right]\right\} .
\end{equation}
Let us go back to equation~(\ref{wf02}) for the coupled oscillators.
If we replace $x_{1}$ and $x_{2}$ by $z$ and $t$ respectively, we
arrive at the above expression for the covariant harmonic oscillators.

We are of course talking about two different physical systems.  For
the case of coupled oscillators, there are two one-dimensional
oscillators.  In the case of covariant harmonic oscillators, there
is one oscillator with two variables.  The Lorentz boost corresponds
to coupling of two oscillators.  With these points in mind, we can
translate the physics of coupled oscillators into the physics of
the covariant harmonic oscillators.

We can obtain the expansion of equation~(\ref{wf03}) from
equation~(\ref{expan1}) by replacing $x$ and $y$ by $z$ and $t$
respectively, and the expression becomes
\begin{equation}\label{expan2}
\psi_{\eta }(z, t) = {1 \over \cosh\eta}\sum^{}_{k}
(\tanh\eta )^{k} \phi_{k}(z) \phi_{k}(t) .
\end{equation}

Thus the space variable $z$ and the time variable $t$ are entangled
in the same manner as given in Ref.~\cite{giedke03}.  However, there
is a very important difference.  The $z$ variable is well defined in
the present form of quantum mechanics, but the time-separation
variable $t$ is not.  First of all, it is different from the calendar
time.  Both father and son become old according to the calendar
time, but their age difference remains invariant.  However, this
time separation becomes different in different Lorentz frames,
because the simultaneity in special relativity is not an invariant
concept~\cite{kn06aip}.

Does this time-separation variable exist when the hadron is at rest?
Yes, according to Einstein.  In the present form of quantum mechanics,
we pretend not to know anything about this variable.  Like the
$y$ variable of section~\ref{restof}, this time-separation variable
belongs to Feynman's rest of the universe.

Using the wave function $\psi_{\eta}(z,t)$ of equation~(\ref{wf03}),
we can construct the pure-state density matrix
\begin{equation}
\rho(z,t;z',t')
= \psi_{\eta}(z,t)\psi_{\eta}(z',t') ,
\end{equation}
which satisfies the condition $\rho^{2} = \rho $:
\begin{equation}
\rho(z,t;z',t') = \int \rho(z,t;z'',t'')
\rho(z'',t'';z',t') dz'' dt'' .
\end{equation}
If we are not able to make observations on $t$, we should take the
trace of the $\rho$ matrix with respect to the $t$ variable.  Then
the resulting density matrix is
\begin{equation} 
\rho(z, z') = \int \rho (z,t;z',t) dt .
\end{equation}

The above density matrix can also be calculated from the expansion
of the wave function given in equation~(\ref {expan2}).  If we perform
the integral of equation~(\ref{integ}), the result becomes
\begin{equation}
\rho(z,z') = \left({1 \over \cosh(\eta)}\right)^{2}
\sum^{}_{k} (\tanh\eta)^{2k}
\phi_{k}(z)\phi^{*}_{k}(z') .
\end{equation}
which becomes identical to the expression of equation~(\ref{dmat}) if
the variables $z$ and $z'$ are replaced by $x$ and $x'$ respectively.
We can then construct the same logic as the one following
equation~(\ref{dmat}) to get the entropy~\cite{hkn89pl}.

Let us summarize.  At this time, the only theoretical tool available
to this time-separation variable is through the space-time
entanglement, which generate entropy coming from the rest of the
universe.  If the time-separation variable is not measured the
entropy is one of the variables to be taken into account in the
Lorentz-covariant system.

In spite of our ignorance about this time-separation variable
from the theoretical point of view, its existence has been proved
beyond any doubt in high-energy laboratories.  We shall see in
section~\ref{feydeco} that it plays a role in producing a decoherence
effect observed universally in high-energy laboratories.

\section{Parton Picture and Decoherence}\label{feydeco}

In a hydrogen atom or a hadron consisting of two quarks, there is a
spacial separation between two constituent elements.  In the case of
the hydrogen atom we call it the Bohr radius.  If the atom or hadron is
at rest, the time-separation variable does not play any visible role
in quantum mechanics.  However, if the system is boosted to the
Lorentz frame which moves with a speed close to that of light, this
time-separation variable becomes as important as the space separation
of the Bohr radius.  Thus, the time-separation plays a visible role
in high-energy physics which studies fast-moving bound states.  Let
us study this problem in more detail.

It is a widely accepted view that hadrons are quantum bound states
of quarks having a localized probability distribution.  As in all
bound-state cases, this localization condition is responsible for
the existence of discrete mass spectra.  The most convincing evidence
for this bound-state picture is the hadronic mass
spectra~\cite{fkr71,knp86}.  However, this picture of bound states
is applicable only to observers in the Lorentz frame in which the
hadron is at rest.  How would the hadrons appear to observers in
other Lorentz frames?

In 1969, Feynman observed that a fast-moving hadron can be regarded
as a collection of many ``partons'' whose properties appear to
be quite different from those of the quarks~\cite{fey69}.  For
example, the number of quarks inside a static proton is three, while
the number of partons in a rapidly moving proton appears to be infinite.
The question then is how the proton looking like a bound state of
quarks to one observer can appear different to an observer in a
different Lorentz frame?  Feynman made the following systematic
observations.

\begin{itemize}

\item[a.]  The picture is valid only for hadrons moving with
   velocity close to that of light.

\item[b.]  The interaction time between the quarks becomes dilated,
    and part  ons behave as free independent particles.

\item[c.]  The momentum distribution of partons becomes widespread as
    the hadron moves fast.

\item[d.]  The number of partons seems to be infinite or much larger
     than that of quarks.

\end{itemize}

\noindent Because the hadron is believed to be a bound state of two
or three quarks, each of the above phenomena appears a
s a paradox,
particularly b) and c) together.  How can a free particle have a
wide-spread momentum distribution?

In order to resolve this paradox, let us construct the
momentum-energy wave function corresponding to equation~(\ref{eta}).
If the quarks have the four-momenta $p_{a}$ and $p_{b}$, we can
construct two independent four-momentum variables~\cite{fkr71}
\begin{equation}
P = p_{a} + p_{b} , \qquad q = \sqrt{2}(p_{a} - p_{b}) .
\end{equation}
The four-momentum $P$ is the total four-momentum and is thus the
hadronic four-momentum.  $q$ measures the four-momentum separation
between the quarks.  Their light-cone variables are
\begin{equation}\label{conju}
q_{u} = (q_{0} - q_{z})/\sqrt{2} ,  \qquad
q_{v} = (q_{0} + q_{z})/\sqrt{2} .
\end{equation}
The resulting momentum-energy wave function is
\begin{equation}\label{phi}
\phi_{\eta }(q_{z},q_{0}) = \left({1 \over \pi }\right)^{1/2}
\exp\left\{-{1\over 2}\left[e^{-2\eta}q_{u}^{2} +
e^{2\eta}q_{v}^{2}\right]\right\} .
\end{equation}
Because we are using here the harmonic oscillator, the mathematical
form of the above momentum-energy wave function is identical to that
of the space-time wave function of equation~(\ref{eta}).  The Lorentz
squeeze properties of these wave functions are also the same.  This
aspect of the squeeze has been exhaustively discussed in the
literature~\cite{knp86,kn77par,kim89}.  The hadronic structure
function calculated from this formalism is in a reasonable agreement
with the experimental data~\cite{hussar81}.

\begin{figure}
\centerline{\includegraphics[scale=0.5]{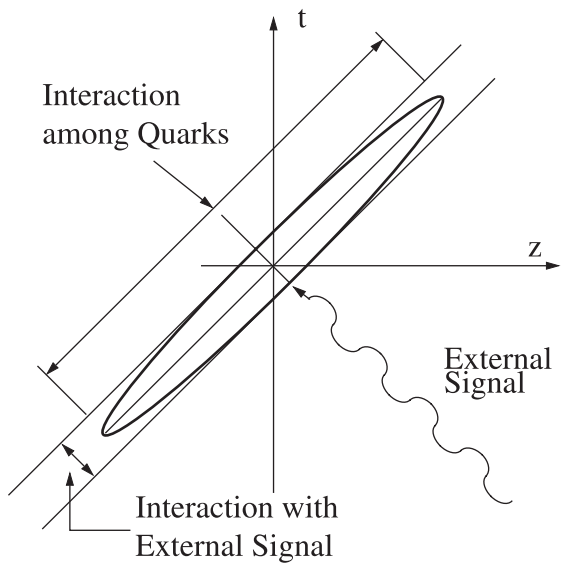}}
\caption{Lorentz-squeezed hadron and interaction times.  Quarks
interact among themselves and with external signal.  The interaction
time of the quarks among themselves become dilated, as the major axis
of this ellipse indicates.  On the other hand, the external signal,
since it is moving in the direction opposite to the direction of the
hadron, travels along the negative light-cone axis.  To the external
signal, if it moves with velocity of light, the hadron appears very
thin, and the quark's interaction time with the external signal
becomes very small.}\label{tdil}
\end{figure}

When the hadron is at rest with $\eta = 0$, both wave functions
behave like those for the static bound state of quarks.  As $\eta$
increases, the wave functions become continuously squeezed until
they become concentrated along their respective positive
light-cone axes.  Let us look at the z-axis projection of the
space-time wave function.  Indeed, the width of the quark distribution
increases as the hadronic speed approaches that of the speed of
light.  The position of each quark appears widespread to the observer
in the laboratory frame, and the quarks appear like free particles.

The momentum-energy wave function is just like the space-time wave
function.  The longitudinal momentum distribution becomes wide-spread
as the hadronic speed approaches the velocity of light.  This is in
contradiction with our expectation from nonrelativistic quantum
mechanics that the width of the momentum distribution is inversely
proportional to that of the position wave function.  Our expectation
is that if the quarks are free, they must have their sharply defined
momenta, not a wide-spread distribution.

However, according to our Lorentz-squeezed space-time and
momentum-energy wave functions, the space-time width and the
momentum-energy width increase in the same direction as the hadron
is boosted.  This is of course an effect of Lorentz covariance.
This indeed is to the resolution of one of the quark-parton
puzzles~\cite{knp86,kn77par,kim89}.

Another puzzling problem in the parton picture is that partons appear
as incoherent particles, while quarks are coherent when the hadron
is at rest.  Does this mean that the coherence is destroyed by the
Lorentz boost?   The answer is NO, and here is the resolution to
this puzzle.

When the hadron is boosted, the hadronic matter becomes squeezed and
becomes concentrated in the elliptic region along the positive
light-cone axis.  The length of the major axis becomes expanded by
$e^{\eta}$, and the minor axis is contracted by $e^{\eta}$.

This means that the interaction time of the quarks among themselves
become dilated.  Because the wave function becomes wide-spread, the
distance between one end of the harmonic oscillator well and the
other end increases.  This effect, first noted by Feynman~\cite{fey69},
is universally observed in high-energy hadronic experiments.  The
period of oscillation is increases like $e^{\eta}$, as indicated in
figure~\ref{tdil}.

On the other hand, the external signal, since it is moving in the
direction opposite to the direction of the hadron travels along
the negative light-cone axis, as is seen in figure~\ref{tdil}.
If the hadron contracts along the negative light-cone axis, the
interaction time decreases by $e^{-\eta}$.  The ratio of the interaction
time to the oscillator period becomes $e^{-2\eta}$.  The energy of each
proton coming out of the Fermilab accelerator is $900 GeV$.  This leads
the ratio to $10^{-6}$.  This is indeed a small number.  The external
signal is not able to sense the interaction of the quarks among
themselves inside the hadron.

Feynman's parton picture is one concrete physical example where the
decoherence effect is observed~\cite{kn05job}.  As for the entropy,
the time-separation variable belongs to the rest of the universe.
Because we are not able to observe this variable, the entropy increases
as the hadron is boosted to exhibit the parton effect.  The
decoherence is thus accompanied by an entropy increase.

Let us go back to the coupled-oscillator system.  The light-cone
variables in equation~(\ref{eta}) correspond to the normal coordinates
in the coupled-oscillator system given in equation~(\ref{normal}).
According to Feynman's parton picture, the decoherence mechanism is
determined by the ratio of widths of the wave function along the two
normal coordinates.

This decoherence mechanism observed in Feynman's parton picture is
quite different from other irreversible decoherences discussed in
the literature.  It is widely understood that the word decoherence
is the loss of coherence within a system.  On the other hand,
Feynman's decoherence discussed in this section comes from the way
external signal interacts with the internal constituents.

\section*{Concluding Remarks}
Modern physics is a physics of harmonic oscillators and/or two-by-two
matrices, since otherwise problems are not soluble.  Thus, all soluble
problems can be combined into one mathematical framework.  Richard
Feynman was always interested in understanding physical problems with
soluble models.  It appears that he had in mind models based on two
coupled harmonic oscillators.

It is now possible to combine Feynman's three papers into one
formalism.  The history of physics tells us that new physics comes
whenever we combine different theories into one.  Maxwell attempted
to combine electricity and magnetism, and he ended up with
electromagnetic waves.  Einstein combined dynamics and electromagnetism
into one transformation law.  He invented relativity.

It is interesting to see that the concept of entanglement is contained
in his rest of the universe.  We see also that the Lorentz-boosted
oscillators and coupled oscillators share the same mathematics, and
that we can learn properties of Lorentz boosts using coupled oscillators
whose physics is very transparent to us.

In addition, using the coupled oscillators, we can clarify the question
of whether the quark model and the parton model are two different
manifestation of one covariant entity.

It would be interesting to combine more of Fyenman's into the
two-oscillator regime which we discussed in the present report.  It
is also possible to include papers written by others in this system.
In fact, in the present report, we included in this regime Dirac's
papers quoted in references~\cite{dir27}-\cite{dir63}.  You may also
consider including your own papers.

\begin{figure}
\centerline{\includegraphics[scale=0.3]{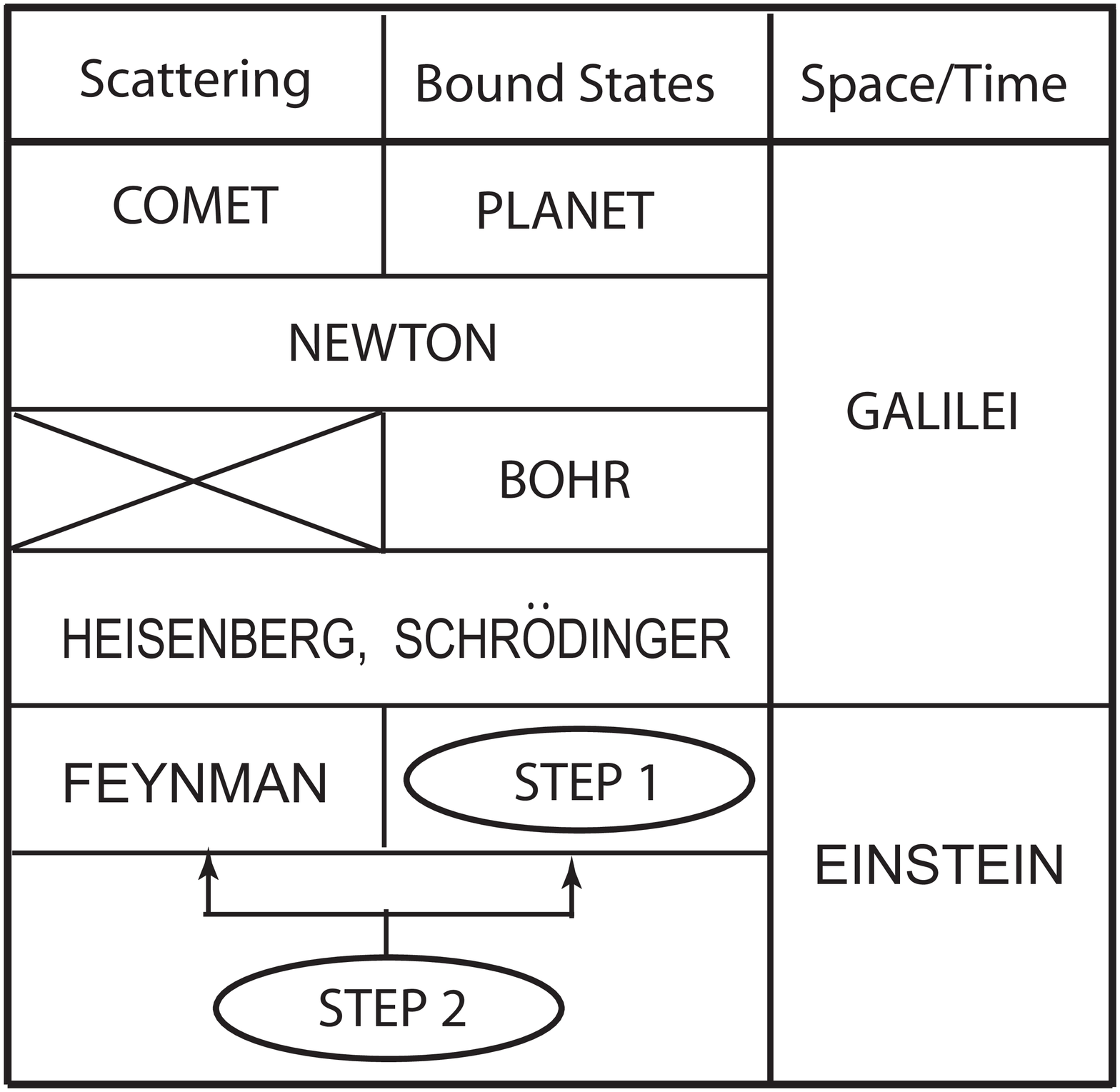}}
\vspace{5mm}
\caption{History of dynamical and kinematical developments.  It is
important to note that mankind's unified understanding of scattering
and bound states has been very brief.  It is therefore not unusual to
expect that separate theoretical models be developed for scattering and
for bound states.  The successes and limitations of the Feynman diagram
are well known.  If we cannot build a covariant quantum mechanics,
it is worthwhile to see whether we can construct a relativistic theory
of bound states to supplement quantum field theory, as Step 1 before
attempting to construct a Lorentz-covariant theory applicable to both
in Step 2.}\label{comet}
\end{figure}

If we assemble those pieces of works done by Feynman and others, we are
able to construct a history table described in figure~\ref{comet}.  As
Feynman stated in 1970, Feynman diagrams are not effective in dealing
with bound-state problems in the Lorentz-covariant regime.  We should
try harmonic oscillators for the bound-state problem.  In this report,
it was shown possible to construct a covariant picture of bound states.

In so doing, we have not introduced any new physical principles.  We
used only the existing rules of quantum mechanics and special relativity,
starting from the quantum mechanics of two coupled oscillators whose
physics is thoroughly transparent to us.

\section*{Acknowledgments}

This report is based mostly on the papers and books I have published
with Marilyn Noz since 1973~\cite{kn73}.  I am deeply grateful to her
for her long-lasting collaboration with me.  Many of the paragraphs of
the present report are directly from the papers which I published with
her.

I am also grateful to Professor Eugene Wigner for spending time with me
from 1985 to 1990.  I used to go to Princeton every two weeks to tell
stories he wanted to hear. In order to make him happy, I had to study
his papers thoroughly, particularly on the Poincar\'e group, space-time
symmetries, group contractions, and on Wigner functions.

During this process, I had to adjust my mode of reasoning to his way of
thinking.  I once asked him whether he thinks like Immanuel Kant.  He
said Yes.  I then asked him whether Einstein was a Kantianist.  He said
firmly Yes, and told me he used to talk to Einstein while they were both
at the Technical University of Berlin.

I asked him, in addition, whether he took a course in philosophy while he
was a student.  He said No.  He became a Kantianist while doing physics.
He then added that philosophers do not dictate people how to think, but
they write down, sometimes systematically, the way in which people think.
Kant was a very systematic person.

Professor Wigner then told me I was the first one to ask whether he was
a Kantianist.  He asked whether I took a course on Kant's philosophy.
I said No.  He then asked me how I came up with Kant's name.  I told
him that Kant is a very popular figure in Japan and Korea, presumably
because Kantianism is very similar to the ancient Chinese philosophy
known as Taoism~\cite{kim95,kim06}.  I told him also that Japan's Hideki
Yukawa's thinking was based on Taoism~\cite{tani79}.

After this conversation, I carried out my systematic study of Immanuel
Kant and his philosophy, including my trip in 2005 to the Russian city
of Kaliningrad, where Kant was born and spent eighty years of his entire
life.  I did my study in the way I do physics, without converting myself
to a philosopher.  Let us assume here that philosophy is a branch of
physics.

If Kant was able to write down his philosophy based on observations he
made in this world, Kaliningrad was his laboratory.  How can one study
Kant without visiting his laboratory?

With this background, I am very happy to say here that Richard Phillip
Feynman was also a Kantianist.  I would like to elaborate this point
in this Appendix.

\begin{appendix}

\section{Feynman's Kantian Inclination}

If Feynman felt that he was moving toward ``One Physics'' while
writing papers on so many different subjects, it is because the
same physics appears to him in different forms.  Feynman had
many different ways of looking at the same thing.  We call this
routinely Kantianism, or the philosophy formulated by Immanuel Kant
(1724-1804).

Why Kant's philosophy so important in physics?.  Is it right to
talk about it in physics papers?  First of all, Einstein's
thinking was profoundly influenced by Kant in his early
years~\cite{howard05,kim95}.  Secondly, we can analyze Kant's
thinking style with the methodology of physics, while avoiding
philosophical jargons.

I came to the United States in 1954 right after my high-school
graduation in Korea.  I did my undergraduate study at the Carnegie
Institute of Technology (now called Carnegie-Mellon University)
and studied at Princeton University for my PhD degree in 1961.
I have been on the physics faculty at the University of Maryland
since 1962.

There is thus every reason to regard myself as an American physicist.
Like all American physicists, I start writing papers if I do not
have ideas.  I gets publishable results while writing.  On the
other hand, I could not get rid of my philosophical background
upon which my brain was configured during my childhood and my
high-school years.  Let us get back to this point in
subsection.~\ref{tao}.

Indeed, because of this background, I was able to raise the question
of whether Feynman was a Kantianist.  In this Appendix, I would like
to explain what Kantianism is in the way physicists explain physics.
I would then like to point out that Feynman was doing his physics in
the way Kant was doing his philosophy.

\subsection{Kantian Influence on Modern Physics}

Unlike classical physics, modern physics depends heavily on observer's
state of mind or environment.  The importance of the observer's
subjective viewpoint was emphasized by Immanuel Kant in his book
entitled "Kritik der reinen Vernunft" whose first and second
editions were published in 1781 and 1787 respectively.

The wave-particle duality in quantum mechanics is a product of
Kantianism.  If your detectors can measure only particle properties,
particles behave like particles.  On the other hand, if your detector
can detect only wave properties, particles behave like waves.
Heisenberg had come up with the uncertainty principle to reconcile
these two different interpretations.  This question is still being
debated, and is a lively issue these days.

Furthermore, observers in different frames see the same physical
system differently.  Kant studied observations from moving frames
extensively.  However, using his own logic, he ended up with a
conclusion that there must be the absolute inertial frame.

Einstein's special relativity was developed along Kant's line
of thinking: things indeed depend on the frame from which
observations are made.  However, there is one big difference.
Instead of the absolute frame, Einstein introduced an extra
dimension, that is the Lorentzian world in which the time variable
is integrated into the three spatial dimensions.

\subsection{Kantianism and Taoism}\label{tao}

I never had any formal education in oriental philosophy, but I
know that my frame of thinking is affected by my Korean background.
One important aspect is that Immanuel Kant's name is known to every
high-school graduate in Korea, while he is unknown to Americans,
particularly to American physicists.  The question then is whether
there is in Eastern culture a ``natural frequency'' which can
resonate with one of the frequencies radiated from Kantianism
developed in Europe.

I would like to answer this question in the following way.  Koreans
absorbed a bulk of Chinese culture during the period of the Tang
dynasty (618-907 AD).  At that time, China was the center of the
world as the United States is today.  This dynasty's intellectual
life was based on Taoism which tells us, among others, that
everything in this universe has to be balanced between its plus
(or bright) side and its minus (or dark) side.  This way of thinking
forces us to look at things from two different or opposite directions.
This aspect of Taoism could constitute a ``natural frequency'' which
can be tuned to the Kantian view of the world where things depend
how they are observed.

I would like to point out that Hideki Yukawa was quite fond of
Taoism and studied systematically the books of Laotse and Chuangtse
who were the founding fathers of Taoism~\cite{tani79}.  Both Laotse
and Chuangtse lived before the time of Confucius.  It is interesting
to note that Kantianism is also popular in Japan, and it is my
assumption that Kant's books were translated into Japanese by Japanese
philosophers first, and Koreans of my father's generation learned
about Kant by reading the translated versions.

In 2005, when I went to Kaliningrad to study the origin of Kant's
philosophy, I visited the Kant Museum twice.  There was a room for
important books written about Kant and his philosophy.  There were
many books written in Russian and in German.  This is understandable
because Kaliningrad is now a Russian city, and the Museum is under
Russian management.  Before 1945, Kaliningrad was a German city
called K\"onigsberg.  Kant was born there and spent eighty years
of his entire life there.  He wrote all of his books in German.

In addition, there are many books written in Japanese.  Perhaps it
could be a surprise to many people, but I was not.  It only
confirmed my Kantian background, as mentioned before.  What is
surprise even to me was that there are no books written in English
in the Kant Museum.  Again this did not surprise me.  Americans
are creative people in the tradition of Thomas Edison.  However,
Edison was not a Kantianist.  I regard myself as an Edisonist and
also as a Kantianist.

Let us see how Taoists can become Kantianists so easily.  For Taoists,
there are always two opposite faces of the same thing called ``yang''
(plus) and ``ying'' (minus).  Finding the harmony between these two
opposite points of view is the ideal way to live in this world.
This is what Taoism is all about.

To Kantianists, however, it is quite natural for the same thing
to appear differently in two different environments.  The problem is
to find the absolute value from these two different faces.  Does this
absolute value exist?  According to Kant, it exists.  To most of us,
it is very difficult to find it if it exists.

Indeed, Kantianism is very similar to Taoism.  It is very easy for
a Taoist to become converted into a Kantianist.  Let us see how
Kant was influenced by the geography of the place where he spent his
entire life.  Let us see also how Taoism was developed in ancient
China.

\subsection{Geographical Origins of Kantianism and Taoism}

Kant was born in the city of K\"onigsberg.  Since he spent eighty
years of his entire life there, his mode of thinking was profoundly
influenced by the lifestyle of K\"onigsberg.  Where is this city?
Like the Mediterranean Ocean, the Baltic Sea was the basin of the
civilization in its own area.  About four hundred years ago,
Lithuania and Poland were very strong countries.  There was a Baltic
costal area between these two countries, which became one of the
commercial centers strong enough to assert independence from its
two strong neighbors.

This place became a country called Prussia and became rich and strong
enough to acquire a large area land west of Poland including Berlin.
Then the center of gravity of Prussia moved to Germany, and the
original Prussia became a province of Germany called "East Prussia."
In Poland, this area is still called Prussia.

K\"onigsberg was a coastal city in East Prussia, but after World War
II, East Prussia became divided into two parts and annexed to Poland
and the Soviet Union.  K\"onisberg is now a Russian city called
Kaliningrad.

Let us look at the map containing both the Baltic and Black Seas.
This area consists of Poland, Belarus, and Ukraine which are between
Eastern and Western Europes.  There are no natural barriers between
this broad borderland,
and anyone with a strong army could walk into or walk through this
area.  Accordingly the city of K\"onigsberg had been under many
different managements~\cite{apple94}.  In addition, since the city
was a commercial center like Venice, there were many visitors with
financial power.  The native citizens of K\"onigsberg therefore had
to entertain those powerful people with many different view points
for the same thing.  Kant wrote philosophy books based on the life
style of his city.

Let us next how Taoism was developed in ancient China.  After the
last ice age, China consisted of many isolated pockets of population.
Those people started moving toward the banks of China's two great
rivers.  This is how today's China started.  When those Chinese came
to the river banks with different languages with different customs,
they were wise enough to realize that they had to live harmoniously
with others.  They started to communicate with others by drawing
pictures, which later became the Chinese characters.  Because they
had different languages, they started singing to convey their
feelings to others.  This is the reason why spoken Chinese still
has tones.

How about different thinkings?  If one has an idea about something,
there were others with different ideas.  They simplified to two
opposite ideas.  If they were to live harmoniously, those two opposite
views should exist together.  They had to develop their philosophy
based on ``Plus and ``Minus,'' and their balance.  This is what the
Taoism is all about.

What is striking is that both Kantianism and Taoism have their
geographical origins.  Both of them are based on the principle of
accommodating different viewpoints.  Modern physics takes into
account the observer's environment and view points.

Modern physics was developed along the line of Kantian philosophy.
If Feynman felt that he was looking for the same thing while writing
papers on different subjects, he was a Kantianist.

\end{appendix}


\begin{thebibliography}{99}

\bibitem{fey69}

Feynman R P 1969 {\em Phys. Rev. Lett.} {\bf 23} 1415 \\
Feynman R P 1969 {\em The Behavior of Hadron Collisions at Extreme
Energies}, in {\em High Energy Collisions}, Proceedings of the
Third International Conference, Stony Brook, New York, edited by
C. N. Yang {\it et al.}, Pages 237-249 (Gordon and Breach,
New York)

\bibitem{fkr71}
Feynman R P, Kislinger M, and Ravndal F 1971 {\em Phys. Rev. D}
{\bf 3} 2706

\bibitem{fey72}
Feynman R P 1972 {\em Statistical Mechanics} (Benjamin/Cummings,
Reading, MA)

\bibitem{dir27}
Dirac P A M 1927 {\em Proc. Roy. Soc. (London)} {\bf A114} 243

\bibitem{dir45}
Dirac P A M 1945 {\em Proc. Roy. Soc. (London)} {\bf A183} 284

\bibitem{dir49}
Dirac P A M 1949 {\em Rev. Mod. Phys.} {\bf 21} 392

\bibitem{dir63}
Dirac P A M 1963 {\em J. Math. Phys.} {\bf 4} 901

\bibitem{howard05}
Howard D A 2005 {\em Physics Today} {\bf 58  No. 12} 34


\bibitem{hkn99ajp}
Han D, Kim Y S and Noz M E 1999 {\em Am. J. Phys.} {\bf 67} 61

\bibitem{kno79ajp}
Kim Y S, Noz M E, and Oh S H, 1979 {\em Am. J. Phys.} {\bf 47} 892

\bibitem{knp86}
Kim Y S and Noz M E 1986 {\em Theory and Applications of the
Poincar\'e Group} (Reidel, Dordrecht)

\bibitem{knp91}
Kim Y S and Noz M E 1991 {\em Phase Space Picture of Quantum Mechanics}
(World Scientific, Singapore)

\bibitem{hkn89pl}
Han D, Kim Y S, and Noz M E 1989 {\em Phys. Lett. A} {\bf 144}
111;

\bibitem{kiwi90pl}
Kim Y S and Wigner E P 1990 {\em Phys. Lett. A} {\bf 147}  343

\bibitem{wig39}
Wigner E 1939 {\em Ann. Math.} {\bf 40} 149

\bibitem{vourdas88}
Vourdas A 1988 {\em Z. Physik} {\bf 71} 527

\bibitem{dodo00}
Dodonov V 2000 {\em J. Opt. B: Quantum Semiclass. Opt.} {\bf 4} R1-R33

\bibitem{yuka53}
H. Yukawa 1953 {\em Phys. Rev.} {\bf 91} 416;
M. Markov M 1956 {\it Nuovo Cimento Suppl.} {\bf 3} 760;
V. L. Ginzburg and V. I. Man'ko 1965 {\em Nucl. Phys.} {\bf 74} 577

\bibitem{giedke03}
Giedke G, Wolf M M, Kr\"uger O, Werner R F, and J. L. Cirac J L
2003 {\em Phys. Rev. Lett} {\bf 91} 107901

\bibitem{kn06aip}
Kim Y S and Noz M E  {\it The Question of Simultaneity in Relativity
and Quantum Mechanics}
{\em Am. Inst. of Phys. Conference Proceedings}
{\bf 0-7354-0301/06} 168

\bibitem{kn77par}
Kim Y S  and Noz M E 1977 {\em Phys. Rev. D} {\bf 15} 335

\bibitem{kim89}
Kim Y S 1989 {\em Phys. Rev. Lett.} {\bf 63} 348

\bibitem{hussar81}
Hussar P E 1981 {\em Phys. Rev. D} {\bf 23} 2781

\bibitem{kn05job}
Kim Y S and Noz M E 2005 {\em J. Opt. B: Quantum and Semiclass. Opt.}
{\bf 7} S458

\bibitem{kn73}
Kim Y S and Noz M E 1973 {\em Phys Rev D} {\bf 8} 3521

\bibitem{kim95}
Kim Y S 1995 {\em Quarks and Partons as Two Different Manifestation of
 One Covariant Entity, in Symmetries in Science VIII} edited by
 B Gruber (Plenum, New York)

\bibitem{kim06}
Kim Y S 2006 {\em Einstein, Kant, and Taoism} Arxiv/physics/0604027

\bibitem{tani79}
Tanikawa Y 1979 {\em Hideki Yukawa: Scientific Works} (Iwanami Shoten,
     Tokyo)

\bibitem{apple94} Applebaum A 1994 {\em Between East and West, Across the
  Borderlands of Europe} (Pantheon Books, New York)

\end{thebibliography}
\end{document}